\documentclass[]{aa} 

\usepackage{graphicx}
\usepackage{txfonts}
\usepackage{xcolor}

\usepackage[breaklinks, colorlinks, citecolor=blue, linkcolor=blue]{hyperref}
\usepackage{listings}

\definecolor{codegreen}{rgb}{0,0.6,0}
\definecolor{codegray}{rgb}{0.5,0.5,0.5}
\definecolor{codepurple}{rgb}{0.58,0,0.82}
\definecolor{backcolour}{rgb}{0.95,0.95,0.92}

\lstdefinestyle{mystyle}{
    backgroundcolor=\color{backcolour},   
    commentstyle=\color{codegreen},
    keywordstyle=\color{magenta},
    numberstyle=\tiny\color{codegray},
    stringstyle=\color{codepurple},
    basicstyle=\ttfamily\scriptsize,
    breakatwhitespace=false,         
    breaklines=true,                 
    captionpos=b,                    
    keepspaces=true,                 
    showspaces=false,                
    showstringspaces=false,
    showtabs=false,                  
    tabsize=2
}

\graphicspath{{/home/aasensio/Dropbox/GIT/DeepLearning/spatially\_variant/validate/figs}}

\DeclareMathOperator*{\argmin}{arg\,min}

\newcommand{\vect}[1]{\boldsymbol{\mathbf{#1}}}
\newcommand{\torchmfbd}{\texttt{torchmfbd}}
\newcommand{\momfbd}{\texttt{MOMFBD}}

\begin{document} 

    \title{\torchmfbd: a flexible multi-object multi-frame\\ blind deconvolution code}
    
    \author{A. Asensio Ramos\inst{1,2}, C. D\'{\i}az Baso\inst{3,4} \and C. Kuckein\inst{1,2} \and S. Esteban Pozuelo\inst{1,2} \and M. G. L\"ofdahl\inst{5}}
    
    \institute{
    Instituto de Astrof\'isica de Canarias (IAC), Avda V\'ia L\'actea S/N,
    38200 La Laguna, Tenerife, Spain\\
    \email{andres.asensio@iac.es}
    \and
    Departamento de Astrof\'isica, Universidad de La Laguna, 38205 La Laguna, Tenerife, Spain
    \and
    Institute of Theoretical Astrophysics,
    University of Oslo, %
    P.O. Box 1029 Blindern, N-0315 Oslo, Norway
    \and
    Rosseland Centre for Solar Physics,
    University of Oslo, %
    P.O. Box 1029 Blindern, N-0315 Oslo, Norway
    \and
    Institute for Solar Physics, Dept. of Astronomy, Stockholm University, AlbaNova University Centre, SE-10691 Stockholm, Sweden
    }

    \date{Received ; accepted }
    
 
    \abstract
    {Post-facto image restoration techniques are essential for improving the quality of ground-based astronomical 
    observations, which are affected by atmospheric turbulence. Multi-object multi-frame blind deconvolution (MOMFBD) methods 
    are widely used in solar physics to achieve diffraction-limited imaging.}
    {We present \torchmfbd, a new open-source code for MOMFBD that leverages the \texttt{PyTorch} 
    library to provide a flexible, 
    GPU-accelerated framework for image restoration. The code is designed to handle spatially variant point spread 
    functions (PSFs) and includes advanced regularization techniques.}
    {The code implements the MOMFBD method using a maximum a-posteriori estimation framework. It supports both 
    wavefront-based and data-driven PSF parameterizations, including a novel experimental approach using 
    non-negative matrix factorization. Regularization techniques, such as smoothness and sparsity constraints, 
    can be incorporated to stabilize the solution. The code also supports dividing large fields of view into patches and 
    includes tools for apodization and destretching. The code architecture is designed to
    become a flexible platform over which new reconstruction and regularization methods can also 
    be implemented straightforwardly.}
    {We demonstrate the capabilities of \torchmfbd\ on real solar observations, showing its ability to produce 
    high-quality reconstructions efficiently. The GPU acceleration significantly reduces computation time, making 
    the code suitable for large datasets. The code is publicly available at \url{https://github.com/aasensio/torchmfbd}.}
    {}
    \keywords{Methods: numerical, data analysis --- techniques: image processing}
    
    \maketitle
    
    \abstract

%
\section{Introduction}
\label{sec:introduction}
The energy injected by the Sun into the Earth's atmosphere is responsible for the
presence of turbulence. Eddies of different scales at different heights in the
atmosphere are responsible for the
creation of refractive index fluctuations that affect the propagation of light. When carrying out
astronomical observations from Earth, these fluctuations produce perturbations in the wavefront, which
produces aberrations both of low order (responsible for tip-tilt, defocus, astigmatism, \ldots) as well
as high order (responsible for a significant decrease in the contrast). 

The turbulence close to the aperture of the telescope (the so-called low-altitude seeing)
is responsible for global aberrations that affect the entire field of view (FoV). These aberrations
are routinely partially corrected at the telescope using adaptive optics (AO) systems that measure the wavefront
distortion and apply a correction to a deformable mirror. If done with sufficient speed, the
correction can significantly improve the image quality. Currently, more than
25\% of the observations in some large-aperture telescopes like Keck and VLT use an AO device
\citep{rigaut15}. Almost all ground-based high resolution solar telescopes are equipped with AO systems so that,
practically, the totality of the solar observations are done with AO. Observations without
AO are only done for specific purposes like off-limb observations, for which measuring the
wavefront is not possible or very challenging. However, even in this case, current advances
look very promising. In fact, solar observatories have pioneered the
development of low-altitude AO systems for visible and near-ultraviolet wavelengths, which are
more challenging than in red or infrared wavelengths. The Swedish 1-m Solar
Telescope \citep[SST;][]{2003SPIE.4853..341S,2024A&A...685A..32S} at the Observatorio del Roque de los Muchachos (Spain) is a case of
success. Newer solar telescopes like the GREGOR telescope \citep{GREGOR, kleint2020} at the Observatorio del Teide (Spain)
or the Goode Solar Telescope (GST) at the Big Bear Observatory (USA) are also equipped with
AO systems.

When turbulence in high-layers is significant, the isoplanatic patch (the region in the sky
where turbulence can be assumed to be spatially homogeneous) becomes smaller, and
it becomes impossible to correct the wavefront for the entire FoV with a single
deformable mirror. In this case, one needs to rely on multi-conjugate adaptive optics (MCAO)
systems, which use several deformable mirrors conjugated to different layers in the atmosphere.
Such systems are still in their infancy in solar observations. 

Even with the best AO systems, the correction is not perfect. The
residual wavefront error is still significant and limits the spatial
resolution and the contrast of the observations. For this reason,
post-facto image restoration techniques are used to push the
observations to the limit and correct the residual wavefront error.
These methods are based on the assumption that the object is constant
across a set of short-exposure images and that the turbulence is frozen
during the exposure time of each image.

The first methods were applied to solar observations before solar
telescopes were equipped with AO. Speckle methods
\citep{labeyrie70,vonderluhe93} are based on knowing the statistics of
the atmospheric turbulence and therefore need a dataset that represents a
statistically relevant sample of the atmosphere, usually $\sim$100 exposures, to
correctly compensate for the attenuation of the power spectrum. For
this reason, speckle methods are less straightforward for data from
modern telescopes as AO correction modifies the statistics. Speckle
methods are still used, e.g., for data from GREGOR and DKIST \citep{2020SoPh..295..172R}, but require
difficult calibrations to produce correct outputs \citep{Fitzgerald_2006}.

Another class of methods is based on fitting a model of the image
formation process to the data. Both the object and the aberrations
from turbulence are estimated jointly by maximizing the likelihood of
the observations under the assumption of Gaussian or Poisson noise.
The first such method was the phase-diversity (PD) technique
\citep{1982OptEn..21..829G,paxman92,paxman92phase}, that makes use of
images that are nominally in focus, together with co-temporal images 
with known added aberrations (commonly just defocus). The PD technique
was developed for solar observations by
\citet{lofdahl_scharmer94}. The more general Multi-Object Multi-frame Blind Deconvolution
technique \citep[MOMFBD;][]{2002SPIE.4792..146L,vannoort05}%
\footnote{In order to avoid confusion between algorithm and
  implementation, we write \momfbd{} when we refer to the code that is widely
  used for solar image restoration with the MOMFBD algorithm. This is the original C++ code by
  \citet{vannoort05} and the version thereof that is maintained by
  T.~Hillberg in Stockholm \citep{2021A&A...653A..68L}.}
can handle a variety of data collection schemes. It uses several
short-exposure images of different objects to produce improved images.
All objects (usually the same solar region observed, e.g., in
different wavelengths within a spectral line) should be observed strictly
simultaneously, so that they are affected by exactly the same
turbulence. PD with two images is used in the data from the
Imaging Magnetometer eXperiment \citep[IMaX;][]{imax11} of the first and second flight of
the Sunrise balloon-borne telescope \citep{sunrise08} or the Tunable Magnetograph \citep[TuMag;][]{2025arXiv250208268D}
of the third flight of Sunrise \citep{2025arXiv250206483K}. MOMFBD is routinely used with and without
PD for the data from the CRisp Imaging Spectropolarimeter
\citep[CRISP;][]{2006A&A...447.1111S, 2008ApJ...689L..69S} and
CHROMospheric Imaging Spectrometer
\citep[CHROMIS;][]{2017psio.confE..85S} of the SST through  its data processing pipeline SSTRED \citep{jaime15,2021A&A...653A..68L}.

Applying MOMFBD to large FoVs is computationally expensive because
of the presence of high-altitude seeing. The wavefront, and the ensuing point spread function (PSF),
changes on scales of a few arcseconds and becomes spatially variant. The computational cost of
applying a per-pixel spatially variant PSF to an image is $\mathcal{O}(N^4)$ (see Sec.
\ref{sec:image_formation}) for an image of size $N \times N$. In contrast, using the Fourier transform to apply a spatially invariant PSF
is only $\mathcal{O}(N^2 \log N)$. Given the prohibitive computational effort required for large 
images, current methods rely on the overlap-add (OLA) method, which requires dividing 
the image in sufficiently small overlapping patches (we often use the term
\emph{patchify} for this purpose), reconstructing each patch 
independently under the assumption of a spatially invariant PSF, and finally merging 
all patches together as a mosaic. The merging needs to be applied with care to avoid artifacts, but the
process can be tuned to produce excellent results.

In this work, we present a new code, \torchmfbd, that implements the MOMFBD method
using the automatic differentiation \texttt{PyTorch} library. The code is designed to be flexible, easy to use, and fast. It is
able to handle spatially variant PSFs using patches or, alternatively, using an experimental method based on
the linear expansion of the PSF in a suitable basis. It also includes several regularization techniques, 
that can be easily switched on and off. Thanks to the use of \texttt{PyTorch}, the code is able to run on GPUs leading
to a significant reduction in the computing time. Additionally, thanks to its modular design,
new reconstruction methods and/or regularization
techniques can be easily implemented. The code is open-source and available online with a very permissive license.

\section{The inverse problem}
\label{sec:inverse_problem}

\subsection{Maximum a-posteriori estimation}
Consider a scenario where $J$ short-exposure observations of $K$ co-spatial objects (e.g., monochromatic images of 
the same region at different wavelengths) are observed simultaneously with 
a ground-based instrument. All observations are, consequently, affected by the presence of turbulence 
in the Earth's atmosphere. If the objects are observed strictly simultaneously, they are affected by exactly the same 
atmospheric disturbances. Additionally, if the exposure time of each observed frame is shorter than the
evolution time of the atmosphere (typically of the order of milliseconds), we can safely assume that the 
atmosphere remains ``frozen'' during this time. 
Furthermore, if the total observation period is shorter than the evolution timescales of the object,
it is reasonable to consider that the object remains unchanged inside the burst of $J$ frames.

From a formal point of view, the observed image $i_{kjp}$ for each object $k=1,\ldots,K$ at frame $j=1,\ldots,J$ and for diversity channel $p=1,\ldots,P$ can be 
approximated using the following generative model for additive noise:
\begin{equation}
\label{eq:generative_model}
i_{kjp} = f(\{o_k, s_{kjp}\}) + n_{kjp},
\end{equation}
where $s_{kjp}$ represents the state of the atmosphere at time $j$ at the wavelength of object $k$ 
and for the diversity channel $p$, 
$o_k$ represents the object $k$ outside of the Earth's
atmosphere and $n_{kjp}$ is a noise term. The function $f(x)$ encodes the image formation, that takes
as input the object and the atmospheric state and produces the observed image. Different 
image formation models will be discussed in the following sections. Following \cite{vannoort05}, we condense
the $j$ and $p$ indices into only one, noting that then $j=1,\ldots,JP$ and that
$s_{kj}$ must include the known diversity for channel $p$.

The solution to the MOMFBD problem boils down to inferring true objects $\vect o=\{o_k\}_{k=1,\ldots,K}$ 
and atmospheric states $\vect s=\{s_{kj}\}_{j=1,\ldots,J;k=1,\ldots,K}$ 
directly from the observed images $i_{kj}$. This is typically achieved by maximizing the 
posterior distribution of these observations under the specific noise assumptions. 
The assumption of uncorrelated Gaussian statistics leads to a Gaussian log-likelihood, so that
the (negative) log-posterior is given by:
\begin{equation}
\label{eq:log_posterior}
L(\vect o,\vect s) = \sum_{k,j,\mathbf{r}} \gamma_{kj} \left\| 
    i_{kj}-f(\{o_k,s_{kj}\}) \right\|^2 + \beta \mathcal{R}(\vect o,\vect s),
\end{equation}
where the summation is carried out over all objects, frames, and pixels (parameterized by the position 
vector $\mathbf{r}$). The maximization of the log-posterior is equivalent to the minimization of the
previous loss function.
This loss function is approximately valid in the solar case due to the large number of photons 
involved and the use of pixellated cameras with negligible cross-talk between pixels. Adding the cross-talk
effect (produced by a potential charge diffusion among pixels) is trivial by including a covariance 
matrix. In cases of low illumination, the multiplicative character
of noise needs to be taken into account, so that the log-likelihood is simply given by a Poisson distribution. 

The term $\mathcal{R}(\vect o,\vect s)$ is a regularization term that comes directly from the log-prior distribution. It 
can be designed to enforce certain properties of the true objects or the instantaneous atmospheric states.
The hyperparameters $\gamma_{kj}$ and $\beta$ need to be tuned for optimal performance. The terms $\gamma_{kj}$ 
represent weights that are often set inversely proportional to the noise variance (images with large amounts 
of noise will contribute less during the optimization). The term $\beta$ is the hyperparameter that controls the 
trade-off between the likelihood and the prior.

\subsection{Image formation}
\label{sec:image_formation}
When dealing with a complex optical system whose PSF varies across space, calculating the intensity 
at each pixel in the image plane involves integrating the product of the PSF at the pixel and the image over the
extension of the PSF. Formally, this can be represented by the following integral equation:
\begin{equation}
\label{eq:general_convolution}
i(\mathbf{r}) = \int o(\mathbf{r}') s(\mathbf{r},\mathbf{r}-\mathbf{r}') d\mathbf{r}'
\end{equation}
where $i(\mathbf{r})$ is the intensity at a given image plane pixel defined by $\mathbf{r} = (x,y)$, 
$o(\mathbf{r}')$ is the object's intensity before interaction with the optical system, evaluated at 
coordinates $\mathbf{r}'=(x',y')$ in the plane of the source, and 
$s(\mathbf{r},\mathbf{r}-\mathbf{r}')$ represents the PSF at location $\mathbf{r}$ as a function of the coordinates
$\mathbf{r}'$ centered at the pixel of interest $\mathbf{r}$. Making explicit the coordinates and
transforming to discrete summations in a pixelated detector of size $N \times N$, Eq.~(\ref{eq:general_convolution}) 
can be written as:
\begin{equation}
  \label{eq:general_convolution_explicit}
i(x, y) = \sum_{x'=1}^{N} \sum_{y'=1}^{N} o(x', y') s(x', y', x-x', y-y').
\end{equation}
Therefore, calculating the intensity at all points requires a computational complexity of $\mathcal{O}(N^4)$, 
as a double summation over the image pixels needs to be carried for all $N \times N$ pixel locations. This
ends up being impractically slow as images grow larger.

\subsubsection{Spatially invariant convolution}
A major simplification occurs when one can assume that the PSF is spatially invariant, so that
it does not change across a certain FoV. Under this condition, 
$s(\mathbf{r},\mathbf{r}-\mathbf{r}')$ does not depend explicitly on $\mathbf{r}$, and
the equality $s(\mathbf{r},\mathbf{r}-\mathbf{r}')=s(\mathbf{r}-\mathbf{r}')$ holds. This allows 
the use of the convolution theorem to compute the intensity at each pixel much more
efficiently:
\begin{equation}
i(\mathbf{r}) = o(\mathbf{r}) \ast s(\mathbf{r}).
\end{equation}
Using this property, the generative model in Eq.~(\ref{eq:generative_model}) can be rewritten as:
\begin{equation}
i_{kj} = o_k \ast s_{kj} + n_{kj}.
\end{equation}
From a computational standpoint, the spatial invariance significantly reduces 
the complexity. The Fast Fourier Transform (FFT) algorithm can be used to compute the
convolution with just three FFT operations. The total computing 
time scales, therefore, as $\mathcal{O}(N^2 \log N)$, due to the logarithmic scaling of FFTs. This 
is the fundamental assumption behind the majority of image deconvolution codes that deal with large images, specifically the \momfbd{} code.
In this case, the loss function can be written as:
\begin{equation}
  \label{eq:log_posterior2}
  L(\vect o,\vect s) = \sum_{k,j,\mathbf{r}} \gamma_{kj} \left\| 
      i_{kj}-o_k \ast s_{kj} \right\|^2 + \beta \mathcal{R}(\vect o,\vect s).
  \end{equation}
Transforming the convolution into the Fourier domain, it can be rewritten as:
\begin{equation}
  \label{eq:log_posterior3}
  L(\vect O,\vect S) = \sum_{k,j,\mathbf{u}} \gamma_{kj} \left\| 
    I_{kj}-O_k S_{kj} \right\|^2 + \beta \mathcal{R}(\vect O,\vect S),
  \end{equation}
where the capital letters denote the Fourier transforms of the corresponding lower-case variables, i.e., $I_{kj}=\mathcal{F}(i_{kj})$,
and the summation over pixels is now carried out in the Fourier plane.
The regularization term can be computed either in the Fourier or the real domain, as needed.

The simultaneous minimization of $L(\vect O,\vect S)$ defined in 
Eq.~(\ref{eq:log_posterior3}) with respect to the object ($\vect O$) and the atmospheric state 
($\vect S$) constitutes a non-convex optimization problem. While gradient-based methods within automatic 
differentiation frameworks such as \texttt{PyTorch} (as employed by \torchmfbd) offer a means 
for the simultaneous optimization in the general case, it is relevant to explore the alternative approach based on alternating 
optimization, also known as block coordinate descent \citep[see, e.g.,][]{10.5555/964885.964886}. This method 
involves iteratively solving the following two optimization sub-problems:
\begin{eqnarray}  
    \vect O^{(k+1)} &=& \argmin_{\vect O} L(\vect O,\vect S^{(k)})\label{eq:subproblem1} \\
    \vect S^{(k+1)} &=& \argmin_{\vect S} L(\vect O^{(k+1)},\vect S).  
\end{eqnarray}
Algorithmically, this entails fixing one set of parameters and optimizing the loss
with respect to the other set of parameters. Specifically, given the current estimate of the atmospheric states 
$\vect S^{(k)}$ at iteration $k$, we optimize for the object parameters $\vect O$, which will
become updated to $\vect O^{(k+1)}$. Subsequently, using 
the updated object parameters $\vect O^{(k+1)}$, we optimize for the atmospheric 
states $\vect S$, that will become $\vect S^{(k+1)}$. This iterative procedure is repeated until a chosen convergence criterion is met.

The alternating optimization scheme is relevant in the specific case of adding no regularization to 
the objects, i.e., $\mathcal{R}(\vect O,\vect S)=\mathcal{R}(\vect S)$. In such case,
the solution to the sub-problem Eq.~(\ref{eq:subproblem1}) can be obtained analytically \citep{paxman92}, and
it is given by a Wiener filter \citep{wiener49}:
\begin{equation}
  \label{eq:object_optimization}
  O_k =  \frac{\sum_j \gamma_{kj} I_{kj} S_{kj}^*}{\sum_j \gamma_{kj} S_{kj} S_{kj}^* + S_n/S_0},
\end{equation}
where $S_n$ is the power spectral density (PSD) of the noise and $S_0$ is an estimation of the PSD of the object 
(the simplest possible Wiener filter assumes $S_n/S_0=C$, with $C$ a constant).
This estimation of the objects can be then plugged back into the loss function which, after some
algebra, can be simplified to:
\begin{equation}
  L(\vect S)=\sum_{k,\mathbf{u}} \left[ \sum_{j} \gamma_{kj} \left|I_{kj}\right|^{2}-   
  \frac{\left|\sum_{j} \gamma_{kj} I_{kj}^* S_{kj}\right|^{2}}
 {\sum_{j} \gamma_{kj} \left|S_{kj}\right|^{2} + S_n/S_0} 
\right].
\label{eq:loss_momfbd}
\end{equation}
Note that this loss function does not explicitly depend on the object. In their PD algorithm, \citet{lofdahl_scharmer94} used a
similar approach, but without the $S_n/S_0$ term in the denominator of
Eq.~(\ref{eq:object_optimization}). They instead regularized the noise by
use of an optimized low-pass Fourier filter (see
Sec.~\ref{sec:fourier_filter}). The equivalent loss function for
multiple frames is used in the MFBD (multi-frame blind deconvolution, equivalent to MOMFBD when $K=1$)
and MOMFBD algorithms of \citet{2002SPIE.4792..146L} and \citet{vannoort05} and
now also in \torchmfbd{}.

\subsubsection{Spatially variant convolution}
\label{sec:spat-vari-conv}
If the PSF is not spatially invariant, the convolution operation needs to be carried out
explicitly. However, huge time savings can 
still be obtained if we make the assumption that the PSF can
be expanded linearly into a set of basis functions from a predefined dictionary:
\begin{equation}
\label{eq:psf_expansion}
s(x', y', x, y) = \sum_{i=1}^M a_i(x', y') \phi_i(x, y).
\end{equation}
The dictionary is composed of the functions $\phi_i(x, y)$ for $i=1,\ldots,M$, which are the same for
all pixels in the image plane. These functions, although they do not need
to be orthogonal, should be efficient at representing the possible PSFs, so that one can set $M$ as low
as possible. The coordinate-dependent coefficients $a_i(x', y')$ encode the amplitude 
of each element of the dictionary for the
spatially variant PSF. 

Due to energy conservation, the PSF needs to integrate to unity
at each location. This can be enforced by making sure the basis functions $\phi_i(x, y)$ integrate
to unity and that $\sum_i a_i=1$ for each pixel $(x',y')$. By introducing this expansion in Eq.~(\ref{eq:general_convolution_explicit}),
one finds:
\begin{equation}
i(x, y) = \sum_{i=1}^M \int_{-\infty}^\infty \int_{-\infty}^\infty o(x', y') a_i(x', y') \phi_i(x-x', y-y') \mathrm{d}x' \mathrm{d}y',
\end{equation}
which can be rewritten as:
\begin{equation}
i(x, y) = \sum_{i=1}^M ( o a_i ) \ast \phi_i(x, y).
\end{equation}
This equation shows that the intensity at each pixel can be computed as a linear expansion, where each term
is the convolution of the product of the object with the coordinate-dependent coefficients $a_i(u, v)$ and the
functions $\phi_i(x, y)$, which are common for all pixels. The computational complexity of this
operation is $\mathcal{O}(N^2 M \log N)$, which is small if $M$ is relatively small. The  
generative model in Eq.~(\ref{eq:generative_model}) can thus be rewritten as:
\begin{equation}
i_{kj} = \sum_{i=1}^M (o_k a_{ikj}) \ast \phi_i + n_{kj},
\end{equation}
where $a_{ikj}$ is the image with the $i$-th expansion coefficients for the $j$-th frame of the $k$-th object. In the case
of a spatially invariant PSF, all $a_{ij}$ are constant and, using
properties of the convolution integral, the generative model turns into a simple weighted sum of the convolution 
of the object with the elements of the dictionary:
\begin{equation}
i_{kj} = \sum_{i=1}^M a_{ikj} (o_k \ast \phi_i) + n_{kj}.
\end{equation}

For the spatially variant PSF, the log-posterior turns out to be written as:
\begin{equation}
  \label{eq:log_posterior4}
  L(\vect o,\vect s) = \sum_{k,j,\mathbf{r}} \gamma_{kj} \left\| 
      i_{kj}-\sum_{i=1}^M (o_k a_{ikj}) \ast \phi_i \right\|^2 + \beta \mathcal{R}(\vect o,\vect s).
  \end{equation}
Transforming the convolution into the Fourier domain, the log-posterior can be rewritten as:
\begin{equation}
  \label{eq:log_posterior5}
  L(\vect O,\vect S) = \sum_{k,j,\mathbf{u}} \gamma_{kj} \left\| 
    I_{kj}-\sum_{i=1}^M \hat O_{ikj} \Phi_i \right\|^2 + \beta \mathcal{R}(\vect O,\vect S),
  \end{equation}
where $\Phi_i$ are the Fourier transforms of the $\phi_i(x, y)$ functions, and $\hat O_{ikj}=\mathcal{F}(o_k a_{ikj})$. 
Since no obvious analytical
solution exists in general for the objects, one needs to optimize the loss function simultaneously 
with respect to the objects and the atmospheric states. On the contrary, in the specific case 
of a spatially invariant PSF, the loss function
can again be solved analytically like in Eq.~(\ref{eq:object_optimization}),
but making the formal substitution $S_{kj} \rightarrow \sum_{i=1}^M a_{ikj} \phi_i$.

\section{PSF parameterization}
We consider two different ways of parameterizing the PSF. The first one is based on the wavefront
at the aperture of the telescope, and the second one is based on an experimental linear expansion of the PSF in a suitable basis.
Both methods have their advantages and disadvantages, and the choice of one or the other depends on the specific
application. 

\subsection{Wavefronts}
The standard approach in solar applications of MOMFBD is to parameterize the PSF 
directly from the wavefront at the aperture of the telescope. This is of special interest in the
case of spatially invariant PSFs. To this end, we define the generalized pupil function $P_{kj}$ for
the observation of frame $j$ and object $k$ as:
\begin{equation}
  P_{kj}(\mathbf{v}) = A_k(\mathbf{v}) \exp \left[ i \varphi_{kj}(\mathbf{v}) \right],
\end{equation}
where $\mathbf{v}$ represents the coordinates on the pupil plane, 
$A_k(\mathbf{v})$ is the mask of the telescope aperture (that can potentially include the 
shadow of a secondary mirror and its mounting on the primary mirror), which we assume to be unity inside
the aperture of the telescope and zero outside, and $i$ refers to the imaginary unit number.
The phase of the generalized pupil function is given by the
wavefront $\varphi_{kj}$, that we make explicit in the following.
The PSF can be obtained from the autocorrelation of the generalized pupil function as follows:
\begin{equation}
s_{kj} = |\mathcal{F}^{-1}(P_{kj})|^2.
\label{eq:pupil_func}
\end{equation}


\begin{figure}
  \centering
  \sidecaption
  \includegraphics[width=\columnwidth]{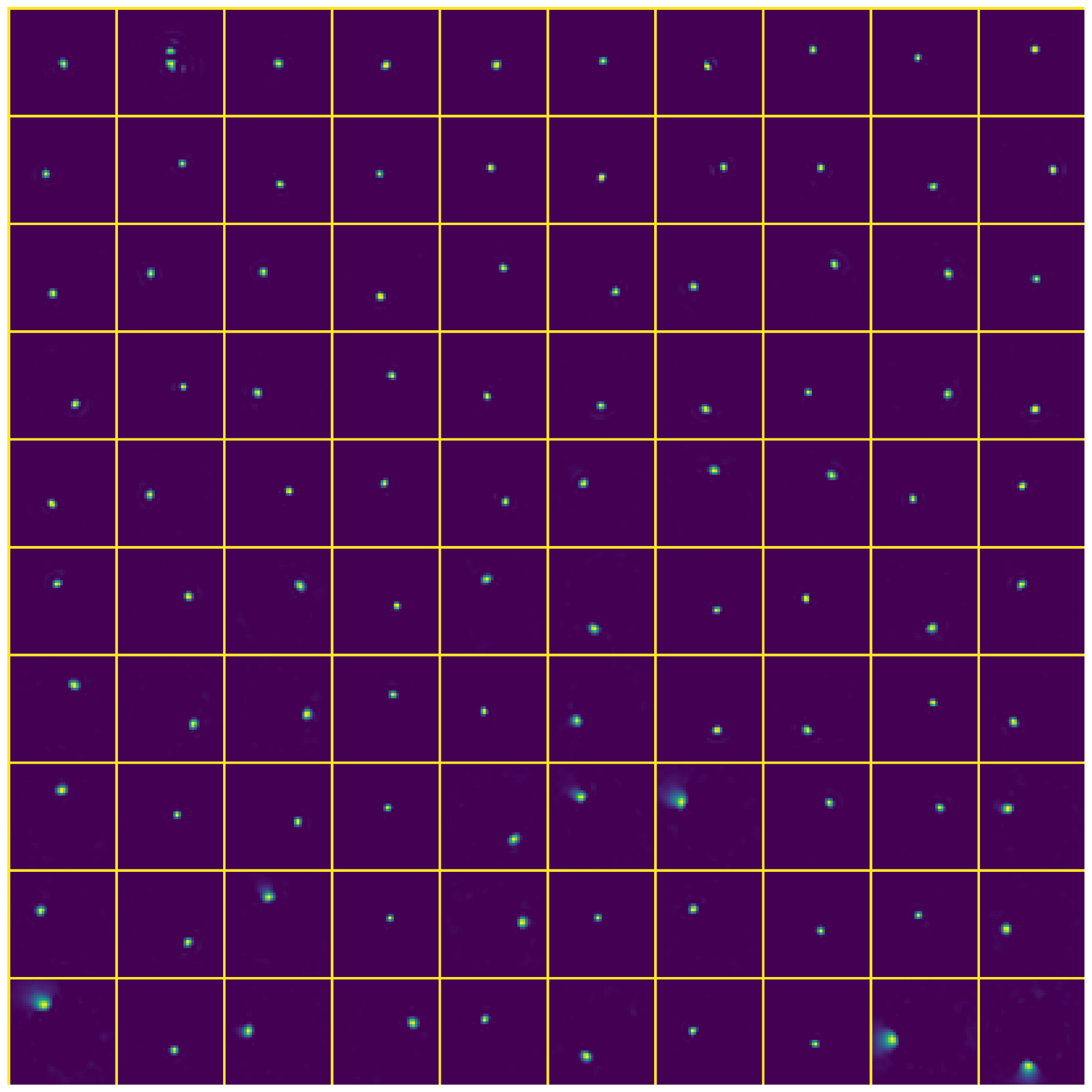}
  \caption{Basis functions of the NMF dictionary for the CRISP instrument at 8542 \AA. The first 100 
  basis functions, out of a total of 150, are shown.}
  \label{fig:basis}
\end{figure}

The wavefronts are typically parameterized through suitable orthogonal basis functions. \torchmfbd\ considers
both the Zernike functions and the Karhunen--Lo\`eve (KL) basis. The KL basis is the optimal set of basis functions for atmospheric turbulence because they are statistically independent for Kolmogorov turbulence. They can be constructed by diagonalization of the Zernike covariance matrix \citep[see][]{roddier90,2021A&A...646A.100A}. 
Zernike modes, although not adapted to atmospheric turbulence, may offer advantages for describing optical aberrations.
The wavefront can be expressed (in this case using the KL modes, but the same
applies with the Zernike modes) as:
\begin{equation}
\varphi_{kj}(v)=\sum_{i=1}^{M}\alpha_{ji}{\rm KL}_i / \lambda_k,
\label{eq:wavefront}
\end{equation}
where $M$ is the number of basis functions and $\alpha_{ji}$ is the coefficient associated with
the basis function $\mathrm{KL}_i$ of the $j$-th observed frame. The normalization by the wavelength ($\lambda_k$)
in the linear expansion is used so the expansion is the same for objects with different wavelengths \citep{vannoort05}.

The expansion in terms of wavefront modes has several interesting properties. First, it is very compact.
Second, PSFs obtained from the wavefront expansion are automatically non-negative and have a compact support,
physical constrains that are important to fulfill.
Finally, low-order modes have well-known physical meaning (tip-tilt, defocus, astigmatism, etc.) and are
mainly responsible for the image quality. For this reason, correcting only a few tens of low-order modes 
produces visually appealing results. High-order modes contribute more to decreasing the contrast of the
observations due to the so-called straylight and correcting for them is more 
challenging. \cite{2010A&A...521A..68S} and \cite{2022A&A...668A.129L}
have demonstrated that compensating for the high-order modes in a statistical sense largely 
removes the dominant source of spatial straylight in SST data.

\begin{figure}
  \centering
  \sidecaption
  \includegraphics[width=\columnwidth]{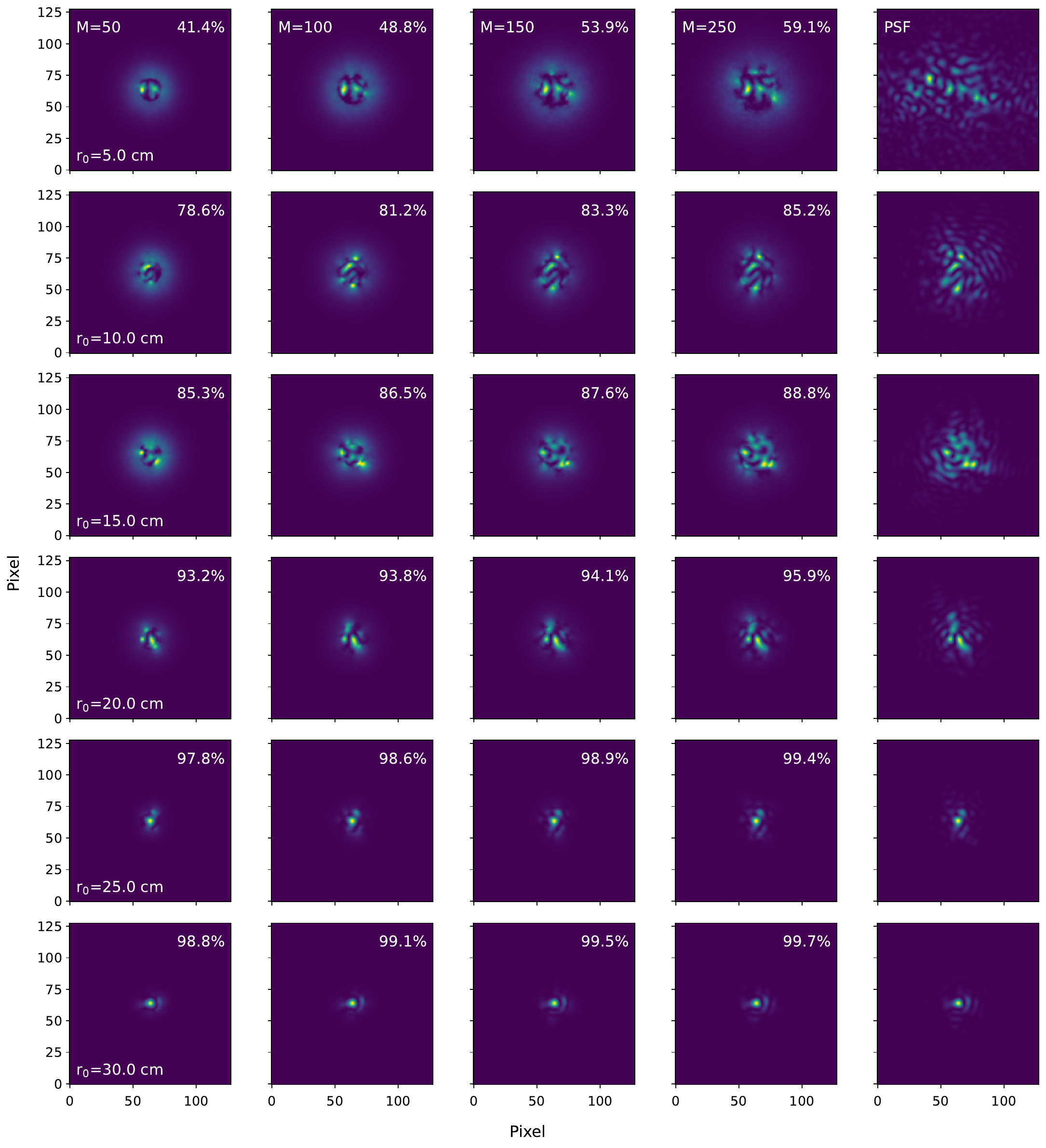}
  \caption{Reconstruction of PSFs with different values of $r_0$ using the NMF dictionary for the 
  CRISP instrument at 8542 \AA. The PSF 
  is shown in the rightmost column, while the NMF reconstructions with different values of $M$ are shown in the other columns.}
  \label{fig:psf_reconstruction_nmf}
\end{figure}

The main disadvantage of the wavefront expansion is that the relation between the wavefront and the PSF is
non-linear, with ambiguities and quasi-ambiguities present in the relation. In particular, the PSF is invariant when all
even  modes of a wavefront change sign. Additionally, indistinguishable PSFs during the 
reconstruction of extended objects can be obtained with very different wavefront coefficients, specially when
high-order modes are involved. As a consequence, 
the wavefront reconstruction is, in general, not a very well-posed problem. The appearance
of ambiguous solutions can be partially solved by using regularization techniques. Some of them
are discussed in Section \ref{sec:regularization}.

\subsection{PSF}
\label{sec:PSF}
Given the ambiguity issues of describing the PSF via the wavefront expansion, we
also consider the direct linear expansion of the PSF as described in Eq. (\ref{eq:psf_expansion}).
This expansion can be used for both spatially invariant and spatially variant PSFs.
It needs to fulfill three key properties. First, it 
should use as few modes $M$ as possible to represent the PSF. Given that the
computational cost of the convolution operation linearly scales with $M$, having a
compact and efficient representation is crucial. Second, the PSF expansion needs
to fulfill the non-negativity of the PSF. Finally, the PSF has to integrate to unity to avoid
energy leakage.

\begin{table}[t]
  \centering
  \caption{Considered instrumental configurations.}
  \label{tab:instruments}
  \begin{tabular}{lcccccc} 
    \hline\noalign{\smallskip}
  Instrument /  & $D$ & $D_2$ & Pixel & $\lambda$ & Label\\
 Telescope  & [m] & [m] & [arcsec] & [\AA] & \\
    \hline\noalign{\smallskip}
  CRISP/SST & 1.0 & - & 0.059 & 8542 & \ion{Ca}{ii} IR\\
  CHROMIS/SST &  1.0 & - & 0.038 & 3934 & \ion{Ca}{ii} K\\
  HiFI/GREGOR &  1.44 & 0.4 & 0.02761 & 4307 & G-band\\
  HiFI/GREGOR &  1.44 & 0.4 & 0.04979 & 7058 & TiO\\
  HiFI/GREGOR &  1.44 & 0.4 & 0.02489 & 3968 & \ion{Ca}{ii} H\\
  IMaX/Sunrise & 1.0 & - & 0.055 & 5250 & \ion{Fe}{i}\\
    \hline
  \end{tabular}
\end{table}

Although analytical linear expansions for the PSF do exist \citep{2010A&A...518A...6A}, 
we have explored the generation of data-driven dictionaries. Data-driven dictionaries
provide a more compact representation of the PSF, thus potentially reducing the value of $M$.
To this end, we generate a database of 5000 PSFs of size 128$\times$128 pixels.
We build the data matrix $\mathbf{X}$ by stacking all PSF images vectorized into vectors
of length $128^2$. The ensuing matrix is of size $n \times m$,
with $n=5000$ and $m=128^2$.
The PSFs depend on the wavelength, pixel size, and specific aperture of the telescope (diameter
of the primary and secondary mirror and, eventually, the presence of a spider), so that
a different dictionary is needed for every combination of telescope and instrument. The configurations
considered in this paper are displayed in Tab.~\ref{tab:instruments}.
The PSFs are obtained by randomly generating wavefronts with Kolmogorov
turbulence according to \cite{roddier90}. The Fried radius ($r_0$) of each PSF is extracted 
randomly between 5 and 30~cm. Values of $r_0$ above 40 cm produce PSFs very close to diffraction, 
at least for the range of diameters considered in this work. Values of $r_0$ below 5 cm are 
discarded because the quality of the observations is very poor. From experience at telescopes
like the SST, no post-facto reconstruction is applied for observations with so small $r_0$.
We have verified that 128$\times$128 pixels is enough for capturing the support
of all considered PSFs. Finally, we have verified that adding tip-tilts to the PSFs produce 
very inefficient dictionaries due to the large displacements. As a consequence, we prefer to
set them to zero during the generation
of the PSFs and take care of tip-tilt during the optimization by locally
shifting the objects using optical flows and bilinear interpolation.


We propose to use the non-negative matrix factorization 
\citep[NMF;][]{1999Natur.401..788L} algorithm\footnote{We use \texttt{scikit-learn} to compute the NMF decomposition.} to compute the dictionary. The resulting 
dictionary is not orthogonal, less efficient in reproducing the PSFs than other options like the
principal component analysis decomposition (PCA) and slightly more time consuming than other decompositions. However, 
it naturally enforces the non-negativity 
of the PSFs, a property that turns out to be crucial. The data matrix $\mathbf{X}$ is decomposed as the 
product of two non-negative matrices:
\begin{equation}
\mathbf{X} \approx \mathbf{H} \mathbf{W},
\end{equation}
where $\mathbf{H}$ is the $n \times M$ matrix with the non-negative expansion coefficients, while $\mathbf{W}$ is the $M \times m$
matrix with the non-negative dictionary. The NMF algorithm requires to define a-priori the number of components 
$M$ to use for the reconstruction. Since the dictionary is non-orthogonal, an arbitrary PSF can then be 
decomposed using the $\mathbf{W}$ dictionary by solving a non-negative least squares problem. Likewise,
a physical PSF can be obtained by linearly combining the columns of $\mathbf{W}$ with non-negative weights. It is also a
well-known fact of the NMF decomposition that the matrices $\mathbf{W}$ and $\mathbf{H}$ are often sparse, with most of the elements
being very close to zero. This can potentially be used in the future to optimize the PSF generation.

Figure \ref{fig:basis} shows the first 100 basis functions of the NMF dictionary
for the CRISP instrument at 8542 \AA. The number of components $M$ in this case is fixed to 150. The NMF dictionary
is very sparse, with most of the basis functions showing a single peak or only a handful of them located at different positions. 
These peaks are used to represent the position of the speckles in the PSFs. Likewise, Fig.~\ref{fig:psf_reconstruction_nmf}
shows the reconstruction of PSFs for the CRISP instrument at 8542 \AA\ with different values of $r_0$ using the NMF 
dictionaries with increasingly larger values of $M$. For large $r_0$, with more compact PSFs, the reconstructions are
excellent, even using $r=50$. For $r_0=5$ cm, the reconstructions suffer even with $r=250$, and only the central part
of the PSF is correctly recovered. We conclude that, for $r_0>15$ cm, the general shape of the PSF is correctly 
recovered with $M$ between 150 and 250. This is quantified in terms of the 
structural similarity index \citep{ssim2004} between the reconstruction and the original PSF.


\section{Regularization}
\label{sec:regularization}
The solution of the blind deconvolution problem is often plagued with artifacts. The information
of high spatial frequencies in the object is partially or totally lost due to the presence of noise in the observations.
Multi-frame methods recover the high spatial frequencies to a limited extent by combining frames that sample
different Fourier frequencies thanks to the variability of the atmospheric conditions and the noise. However,
regularization techniques are often required to stabilize the solution and reduce the
presence of artifacts. \torchmfbd\ includes flexible interfaces to use one or many of the regularization 
techniques described in the following or to code new ones. 

Regularization methods are split into two
categories: implicit and explicit regularization. Implicit regularization techniques are those
that are applied during the optimization process and cannot be translated 
into a regularization term $\mathcal{R}$ in the loss function. Explicit regularization techniques
appear directly as a penalty term $\mathcal{R}$.

\subsection{Implicit regularizations}
\subsubsection{Fourier filtering}
\label{sec:fourier_filter}
One of the most useful implicit regularization methods is the multiplication of the estimation of the object 
at each iterative step with a low-pass filter in the Fourier domain:
\begin{equation}
O^*_k = H_k \cdot O_k.
\end{equation}
Two options are currently available in \torchmfbd. The first one is a top-hat filter
that sets $H=0$ for frequencies (in units of the diffraction limit of the telescope)
above a certain upper cutoff ($f_u$) and $H=1$ for frequencies below a lower cutoff ($f_l$). 
Between $f_l$ and $f_u$, the filter introduces a smooth transition with a cosine function. Typical
values giving good results lie around $f_l=0.7$--$0.9$, with $f_u \approx f_l+0.05$, 
although this critically depends on the
noise properties of the observations and the amount of observed frames.

A second more refined version of the Fourier filter is the one introduced by \cite{lofdahl_scharmer94},
which can only be applied in the case of a spatially invariant PSF because it is computed with the
following expression:
\begin{equation}
\label{eq:filter_fourier}
  H_k = 1 - \frac{\sum_j \left|S_{kj} \right|^2 }{\left|\sum_j I_{kj} S_{kj}\right|^2},
\end{equation}
which depends on the PSFs and the observed frames. Following the suggestions proposed 
by \citet{lofdahl_scharmer94}, all locations with $H_k<0.2$ are set to zero and all values with $H_k>1$ are set to one. 
As a final step, the filter is smoothed with a median filter with a window of $3 \times 3$ and
all remaining peaks that are not connected with the peak at zero frequency are annihilated.

\subsubsection{Complexity scheduling}
The optimization of the loss function can be very challenging when the PSF is very complex.
It is a good practice to start the optimization with a very simple PSF, and then slowly increase 
the complexity of the PSF. We do this by using an annealing 
schedule in the optimization process, which adds more modes to the wavefront or PSF expansions as the optimization
progresses. In the case of the wavefront expansion, we always start with tip and tilt modes, and 
then gradually increase the complexity. The schedule can follow a linear or sigmoidal increase. We have 
verified that this scheduling is very useful to avoid local minima and to stabilize the optimization process.

\subsection{Explicit regularizations}
\label{sec:explicit}
\subsubsection{Smoothness}
\label{sec:smoothness}
When deconvolving noisy images, the inferred object tends to contain spatially correlated noise. One
way of reducing the amount of noise is by explicitly imposing smoothness on the object
using a Tikhonov regularization (also known as $\ell_2$ regularization) for the spatial gradients:
\begin{equation}
  \label{eq:regularization_smoothness}
  \mathcal{R}_o(\vect o,\vect s) = \lambda_o \sum_k \left\| \nabla o_k \right\|^2. 
\end{equation}
In the case of the spatially variant PSF, the smoothness can also be imposed on the PSFs:
\begin{equation}
\mathcal{R}_s(\vect o,\vect s) = \lambda_s \sum_{i,k,j} \left\| \nabla a_{ikj} \right\|^2.
\end{equation}
This formulation involves the introduction of new hyperparameters $\lambda_o$ and $\lambda_s$, 
which need to be tuned for optimal performance.

\subsubsection{Sparsity}
\label{sec:sparsity}
Another successful and recent regularization technique is the imposition of sparsity on the 
object \citep[see, e.g.,][]{candes06}. This can be done
by using the $\ell_1$ norm of the object in an appropriate basis:
\begin{equation}
  \mathcal{R}_o(\vect o,\vect s) = \lambda_w \sum_k \left\| \mathbf{T}(o_k) \right\|_1,
\end{equation}
where $\mathbf{T}$ is a transformation that maps the object into a sparse domain. The most common
choice is the wavelet transform, which is known to produce sparse representations of the object.
Currently, the code offers the use of the isotropic undecimated wavelet 
transform \citep[IUWT;][]{1994A&A...288..342S,2007ITIP...16..297S}, although adding other
transforms is straightforward and will be done in subsequent versions of \torchmfbd.
The IUWT is a redundant wavelet transform that is computed by applying a series of low-pass and high-pass 
filters to the image, producing a multiscale representation of the image. The IUWT is invertible, 
so that the original image can be reconstructed
from the wavelet coefficients. The hyperparameter $\lambda_w$ needs to be tuned to obtain the best results.
Other transforms

\begin{figure}
    \centering
    \sidecaption
    \includegraphics[width=\columnwidth]{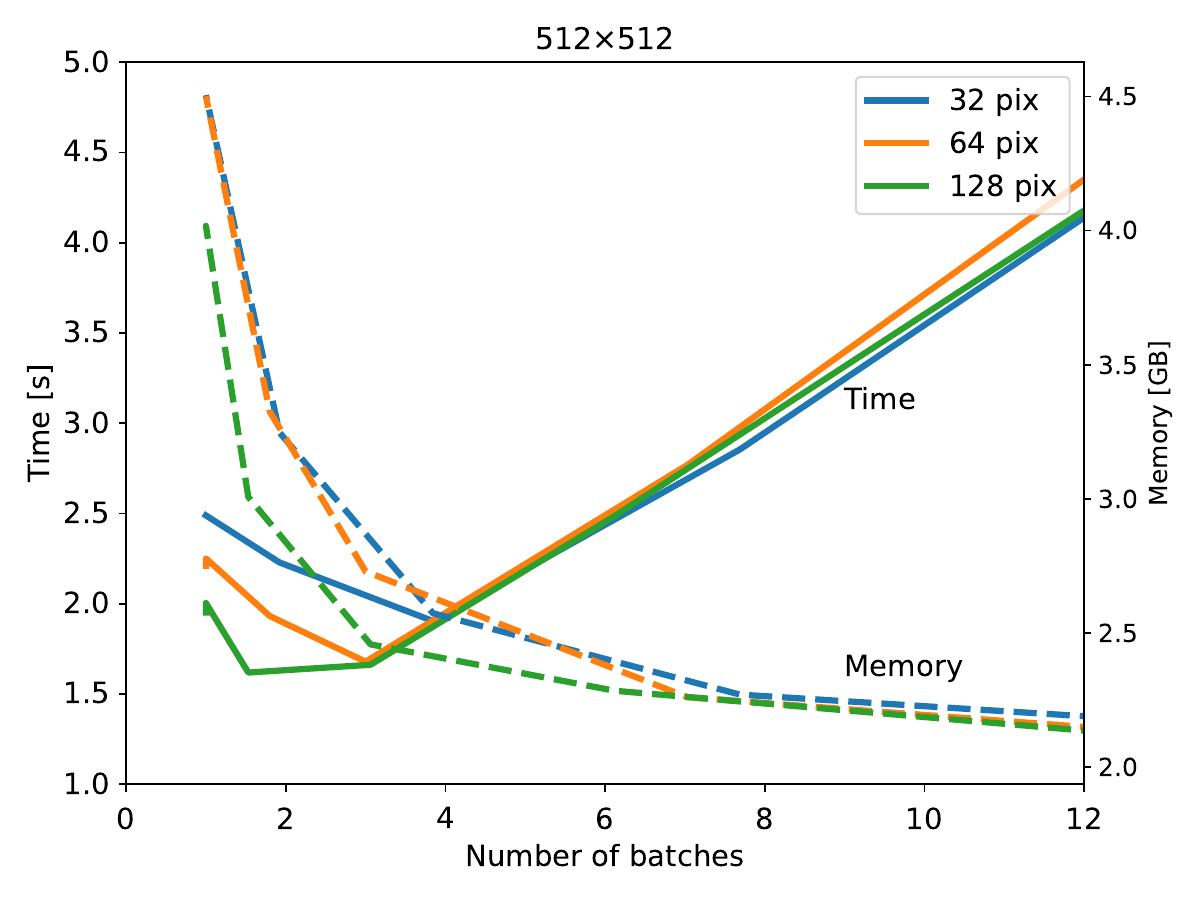}
    \caption{Computing time (solid lines) and memory consumption (dashed lines) for reconstructing a pair of WB and NB images with 15 frames
    in an NVIDIA RTX 4090 GPU as a function of the number of batches. Patches of different size are considered with different colors.}
    \label{fig:timing}
  \end{figure}

\section{\torchmfbd}
\subsection{Destretching}
\label{sec:destretching}
High-altitude turbulence can produce large tip-tilts that vary across the FoV and change
from one frame to the next. We reduce this effect by first producing a destretched version of the
observed frames, in which local tip-tilts are strongly dampened. This is particularly useful
for the later optimization process. We find this strategy very useful and more straightforward than the
iterative estimation of the tip-tilt carried out in the \momfbd{} code, as explained in \cite{vannoort05}.

The destretching proceeds by selecting a frame as a reference and 
estimating the optical flow (essentially the local tip-tilts) between the reference 
frame and the rest of the frames. The estimation is carried out by optimizing a loss function computed
from the modified correlation coefficient of \cite{Evangelidis08} using the Adam optimizer \citep{adam14}. In order to reduce
the number of unknowns, the optical flow can be defined in a coarse grid of $N_\mathrm{opt} \times N_\mathrm{opt}$ pixels
(which can potentially be equal to the size of the frames) and then bilinearly interpolated (if needed) to the original 
frame size. The correlation coefficient is supplemented with a regularization term that penalizes non-smooth 
estimations of the tip-tilt, similar to those described in Eq.~(\ref{eq:regularization_smoothness}). 
Destretching is not perfect, but any remaining tip-tilt is corrected later during the reconstruction process.

\subsection{Division in patches and apodization}
For cases in which the isoplanatic patch is smaller than the FoV, the observations need to be divided 
in patches so that the PSF can be assumed to be constant in each one. Deconvolved patches are later 
merged together or mosaicked to produce the final image. To this end, \torchmfbd\ provides tools to 
easily carry out the division into patches and merging them together to form a mosaicked image. 
One only needs to select the size of the patch 
and the overlap between patches and \torchmfbd\ will compute the number of
patches automatically. The merging is carried out by averaging all patches taking into account
the overlap and using different weighting schemes. The simplest one is to average all patches 
with the same weight, 
which produces some artifacts in some cases. A more sophisticated approach is to use a 
Gaussian weighting starting
from the center of each patch with a user-provided standard deviation. In our experience, the best 
visual results are obtained
using a cosine apodization that smoothly reduces the weight of the patches towards the border 
with adjustable transitions.

\begin{figure}  
  \begin{lstlisting}[label={lst:yaml},caption={Example of the YAML configuration file. In this case, two objects (wideband and narrowband images) observed with CRISP
    are reconstructed. No regularization is used.}]
  telescope:
      diameter: 100.0
      central_obscuration : 0.0
      spider : 0
  images:
      n_pixel : 64    
      pix_size : 0.059
      apodization_border : 6
      remove_gradient_apodization : False
  object1:
      wavelength : 8542.0
      image_filter: scharmer
      cutoff : [0.75, 0.80]  
  object2:
      wavelength : 8542.0
      image_filter: scharmer
      cutoff : [0.75, 0.80]      
  optimization:
      gpu : 0
      transform : softplus
      softplus_scale : 1.0    
      lr_obj : 0.02
      lr_modes : 0.08  
  regularization:
       iuwt1:
          variable : object
          lambda : 0.00
          nbands : 5  
  psf:
      model : kl
      nmax_modes : 44  
  initialization:
      object : contrast
      modes_std : 0.0  
  annealing:
      type: linear
      start_pct : 0.1
      end_pct : 0.6
  \end{lstlisting}  
  \end{figure}  

Given that solar images and/or patches are not periodic in general, apodization is needed to compute all Fourier
transforms and reduce the appearance of wrap-around artifacts. We use a modified 
Hanning window\footnote{A tophat window with a value of unity and a cosine transition
to zero on the borders.} to apodize the patches, which turns out to be a good compromise between reducing the
artifacts produced by the Fourier transform and preserving the information in the images. The width of the apodization
border is set by the user. Before applying the apodization window, we subtract the mean of the 
patch, which is then added back after the apodization. A more sophisticated approach, that is
available in \momfbd{}, consists of subtracting a linear gradient in the image \citep{1987A&A...177..265V}. This reduces
the wrap-around artifacts when strong brightness gradients are present, of special relevance close to active regions.
\torchmfbd\ also includes this option. In case patches are used, the overlapping is suggested to be larger
than twice the apodization border to reduce its impact on the final reconstruction.

\subsection{Optimization and paralellization}
\label{sec:optim-paral}
The optimization of the loss function proceeds by using the Adam optimizer \citep{adam14} and
making use of the automatic differentiation capabilities of \texttt{PyTorch} \citep{pytorch19}. Other
optimizers can be used in \torchmfbd\ straightforwardly.
If available,
all operations are offloaded to a GPU, which can speed up the optimization process by a large margin.
When optimizing the loss functions of Eq.~(\ref{eq:log_posterior3}) or Eq.~(\ref{eq:loss_momfbd}), the calculation
requires $K \times J$ inverse FFTs for the computation of the autocorrelation of the generalized pupil function and additional
$K \times J$ FFTs for the computation of the Fourier transform of the PSFs. When using the NMF dictionary, we remove the
need for the calculation of the autocorrelation and only $K \times J$ FFTs are required. 

\begin{figure}    
    \begin{lstlisting}[language=Python,label={lst:python},caption={Example of a Python script to reconstruct two objects observed with CRISP.}]
    import torch
    import torchmfbd
    # Frames is an array of shape
    # (n_scans, n_obj, n_frames, nx, ny)
  
    # Transform into tensor
    frames = torch.tensor(frames.astype('float32'))
    
    # Object to do patches
    patchify = torchmfbd.Patchify4D()    
            
    # Deconvolution object  
    dec = torchmfbd.Deconvolution('kl.yaml')
    
    # Patchify and add the frames of the two objects
    for i in range(2):        
        patches = patchify.patchify(frames[:, i, :, :, :], 
                                    patch_size=64, 
                                    stride_size=50)
        dec.add_frames(patches, 
                         id_object=i, 
                         id_diversity=0, 
                         diversity=0.0)
    
    # Carry out the deconvolution         
    dec.deconvolve(infer_object=False, 
                      optimizer='adam', 
                      simultaneous_sequences=90,
                      n_iterations=150)
    
    # Extract the deconvolved objects merging the patches
    obj = []
    for i in range(2):
        tmp = patchify.unpatchify(dec.obj[i], 
                                  apodization=6, 
                                  weight_type='cosine', 
                                  weight_params=30)
        obj.append(tmp.cpu().numpy())
    \end{lstlisting}
    \end{figure}

The Adam optimizer requires the tuning of the learning rate. After some experimentation, we have found that
a learning rate of 0.08 for the modes produces good results. If the object is also optimized 
simultaneously, a smaller learning rate of 0.02 is recommended. 
Taking advantage of the parallelization capabilities of \texttt{PyTorch}, the optimization of many patches can be 
carried out in parallel. The loss of all patches is summed up and the optimizer is updated with the total loss.
Since the inferred mode coefficients are not shared between patches, the optimization is, in practical terms, 
independent for each patch.
The number of patches to consider simultaneously is a trade-off between the available memory
and the computational resources. For GPUs with enough memory and depending on the size of the patches, 
the reconstruction of a whole image can be carried out in a single step.

The current version of \torchmfbd\ can easily reconstruct burst of images taken at several
objects with or without a phase diversity channel. The current limitation is that the number
of images in the burst has to be the same for all objects. In order to allow for 
a different number of images (and other complex configurations), we plan to implement the 
linear equality constraints approach 
of \cite{2002SPIE.4792..146L}. Several approaches will be explored for this purpose, each
one with advantages and disadvantages. The first one is the method used in \momfbd{}, which 
is described in \cite{2002SPIE.4792..146L}. The second one is based on adding a penalty term,
similar to those described in Sec. \ref{sec:explicit}, that enforces the approximate fulfillment of the constraints.
A final approach is based on the iterative application of proximal projection
algorithms \citep{parikh2013proximal}, similar to those considered in \cite{2015A&A...577A.140A}.

\begin{figure*}
  \centering
  \sidecaption
  \includegraphics[width=\textwidth]{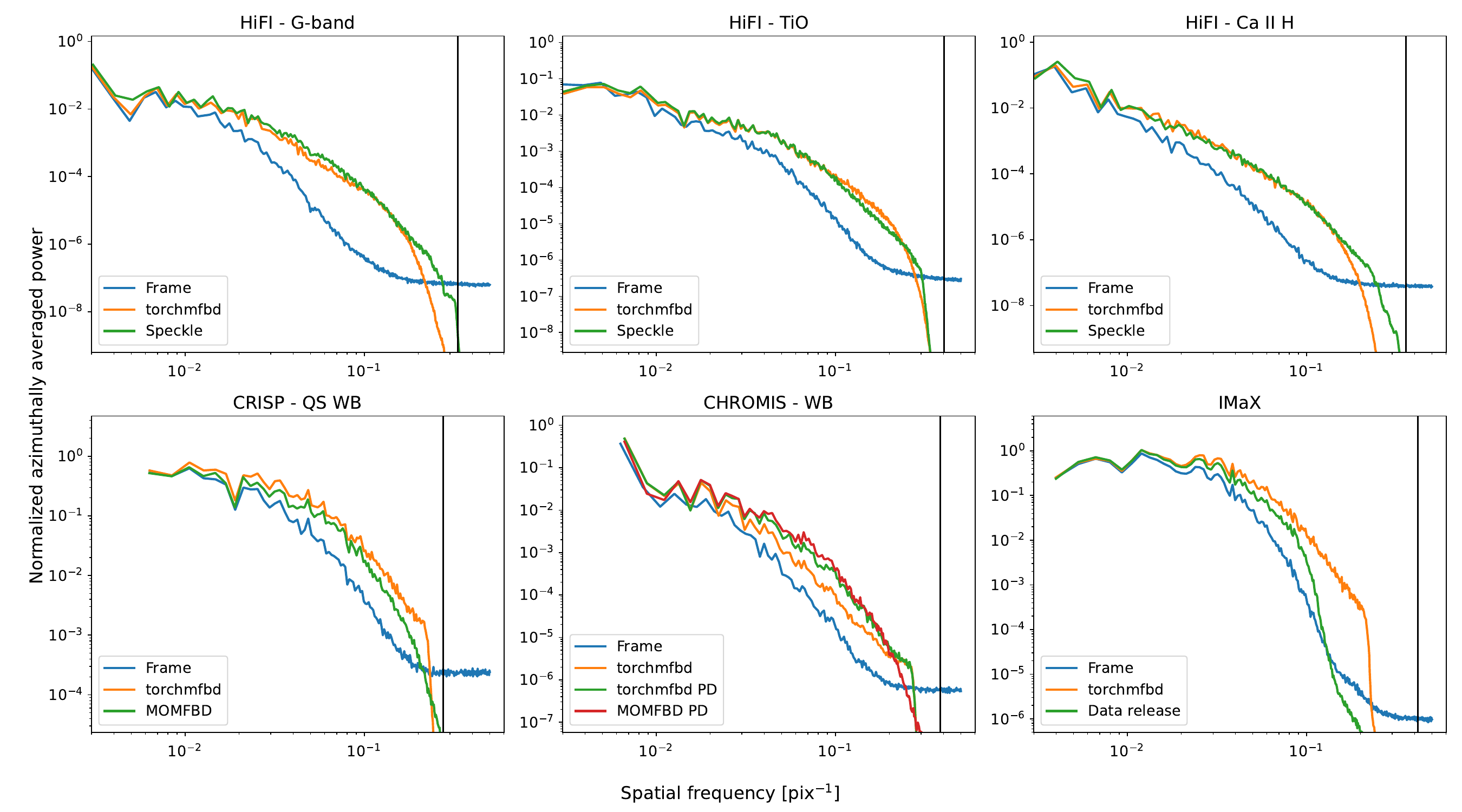}  
  \caption{Azimuthally averaged power spectra for relevant observations shown in Sec.~\ref{sec:results}. The
  average is only performed for angles in the Fourier plane with $-45\pm 15^\circ$ and $45\pm 15^\circ$.  
  The vertical line marks the diffraction limit. Note that the power spectrum is computed in the full FoV, so that
  some high frequency artifacts can appear when the FoV is built from mosaicked patches. To avoid
  showing noise-dominated artifacts, the power below an estimated noise floor (noise level per 
  frame divided by the number of frames) is not displayed.}
  \label{fig:power}
\end{figure*}

\subsection{Timing and memory consumption}
\label{sec:timing-memory-cons}
Although calculations can be performed in CPUs, arguably one of the most interesting advantages of \torchmfbd\ 
is the ability to leverage GPUs to accelerate the calculations. As a consequence, a standard desktop or laptop
with a powerful GPU is enough to deal with large datasets. The results of the scaling is shown 
in Fig.~\ref{fig:timing}, where we measure the computing time and memory consumption for reconstructing 
the sunspot data discussed later in Sect.~\ref{sec:CRISP}. A burst of 15 images of the FoV of 
512$\times$512 pixels is observed simultaneously in a wideband (WB) and narrowband (NB) channels. For
reconstruction, the images are divided in patches of size 32$\times$32, 64$\times$64 and 128$\times$128 pixels, resulting
in a total of 961, 225 and 49 patches, respectively. Taking advantage of the parallelization capabilities of GPUs,
we carry out the optimization of all patches in batches. Figure~\ref{fig:timing} displays the computing time as a 
function of the number of batches required to finalize all patches. The fastest computing times occur
when the batch size is tuned so that the whole FoV is done using $1-4$ batches. For instance, 
when patches of $64\times$64 pixels are used, the optimal computing time is found for a batch size of 75 patches. If smaller 
batches are used (thus requiring more batches), the parallelization capabilities of GPUs are not used 
efficiently and the computing time goes up.
In terms of memory consumption, it goes up when using a smaller number of
batches (equivalently, larger patches), but a good compromise is also found
in the range of $1-4$ batches per FoV.

\subsection{Configuration and execution}
The code is configured using a YAML\footnote{\url{https://yaml.org}} human-readable file that contains 
all the parameters needed to run the code. An example of a configuration file is shown in Listing \ref{lst:yaml}. All 
parameters are self explanatory and, although their description and options can be consulted in the
documentation\footnote{\url{https://aasensio.github.io/torchmfbd}}, we briefly describe them here:
\begin{itemize}
\item \texttt{telescope}: contains the diameter, the central obscuration of the telescope, and the size of
spider. This is used to define the entrance pupil.
\item \texttt{images}: contains the number of pixels of images or patches, the pixel size in arcsec, and the apodization border in pixel.
\item \texttt{object1}, \texttt{object2}, \ldots: contains the wavelength of the object, the Fourier filter used, and the cutoffs of the filter.
\item \texttt{optimization}: contains the index of the GPU to use, the transformation and the scale to use for the optimization of 
the object, and the learning rates.
\item \texttt{regularization}: contains the regularization techniques to use and their hyperparameters.
\item \texttt{psf}: contains the model of the PSF and the number of modes to use.
\item \texttt{initialization}: contains the type of initialization of the object and the standard deviation of the modes (only valid 
if the object is optimized).
\item \texttt{annealing}: contains the type of annealing to use and the start and end percentages of the annealing.
\end{itemize}

The reconstruction can be performed by instantiating a \texttt{torchmfbd.Deconvolution} object, adding the observations
indicating the object and the diversity channel (and the diversity, currently only defocus), and calling 
the \texttt{deconvolve} method of the object. An example is found in Listing \ref{lst:python}.

\section{Results}
\label{sec:results}
We demonstrate the capabilities of \torchmfbd\ with examples from different telescopes. The properties of the
observations are summarized in Table~\ref{tab:instruments}. In this table, $D$ refers to the diameter
of the primary entrance pupil (be it a mirror or lens), and $D_2$ the diameter of a secondary mirror when it casts a shadow on the primary.

As part of the results, we also discuss the power spectra of the
restored images and compare different methods. The power spectra
are shown together in Fig.~\ref{fig:power}. Calculating 1D power spectra
of 2D images, where the scene continues beyond the FoV, is in general
not as straightforward as averaging over all angles. A well-known
complication is that digital Fourier transforms are based on an
assumption of periodic functions. For regular FFT this means the FoV
is implicitly repeated in all directions. The resulting
discontinuities at the border of the FoV cause power artifacts along
the axes in the Fourier domain. These artifacts can be
reduced by apodization but are difficult to avoid completely. Our 1D
power spectra are therefore averages over 30\degr{} wide sectors along
the $\pm45\degr$ directions.

\begin{figure}
  \centering
  \sidecaption
  \includegraphics[width=\columnwidth]{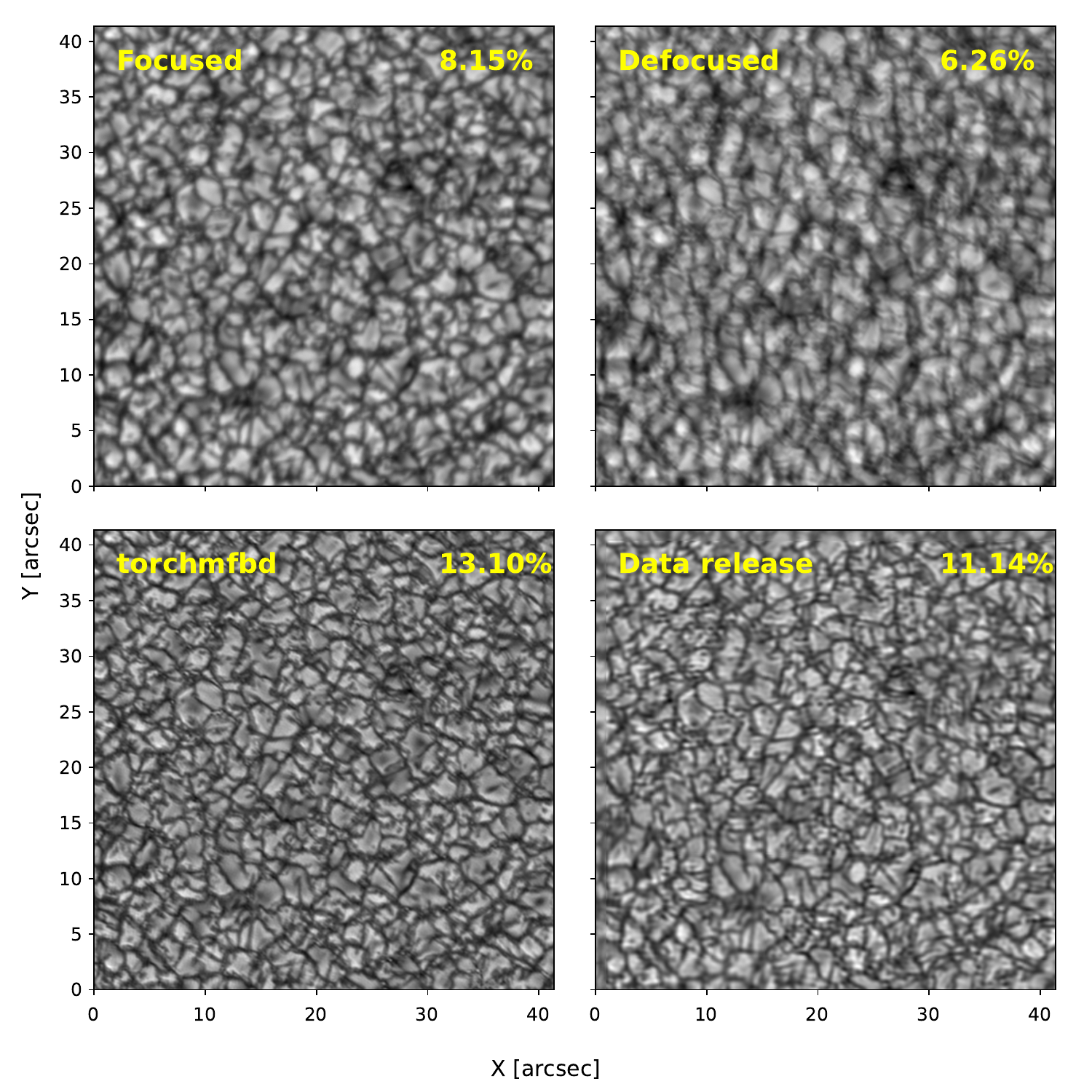}
  \caption{Results of the reconstruction of IMaX data using single-frame phase diversity.
  The upper panels show the focused and defocused frames. The lower panels display
  the results from \torchmfbd\, alongside the reconstructed image of the
  original data release.}
  \label{fig:imax}
\end{figure}

Less well-known is the fact that the mosaicking of independently
restored subfields causes artifacts in the 2D power spectra at high
spatial frequencies in all directions. We therefore interpret the
power spectra with caution in the high spatial frequency regime where
the data are not completely signal-dominated. Notably, the restored
IMaX images are an exception as they are not mosaics, the entire FoV 
shown in Fig.~\ref{fig:imax} was deconvolved as a whole.

The noise power in a multi-frame deconvolved image decreases with the
number of frames. The lower limit of the vertical axis in our plots is
set accordingly. 

\begin{figure*}
    \centering
    \sidecaption
    \includegraphics[width=\textwidth]{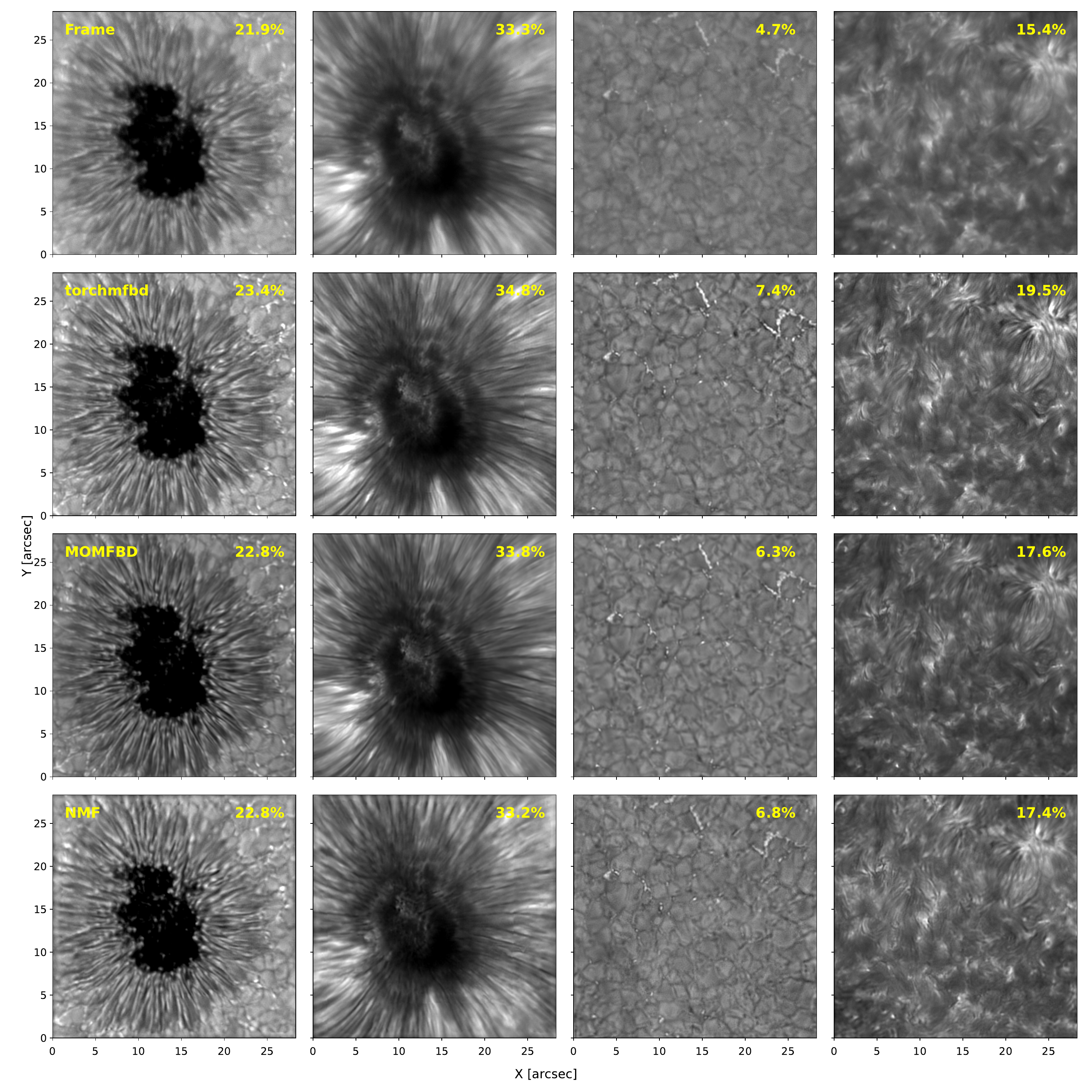}
    \caption{Examples of image reconstructions for data acquired in the Ca \textsc{ii} 8542 \AA\ line with 
    CRISP/SST. The first row displays the
    best frame of the observations in terms of contrast, both for the wideband (first and third columns)
    and narrowband channels (second and fourth columns) for two different FoV. The second,
    third, and fourth rows show the reconstructed images using \torchmfbd, \momfbd{}, and the
    experimental NMF parameterization, respectively. Hyperparameters have been slightly tuned for each observation.
    The contrast displayed in each panel has been computed over the entire image.}
    \label{fig:crisp}
  \end{figure*}

\begin{figure*}
  \centering
  \sidecaption
  \includegraphics[width=0.7\textwidth]{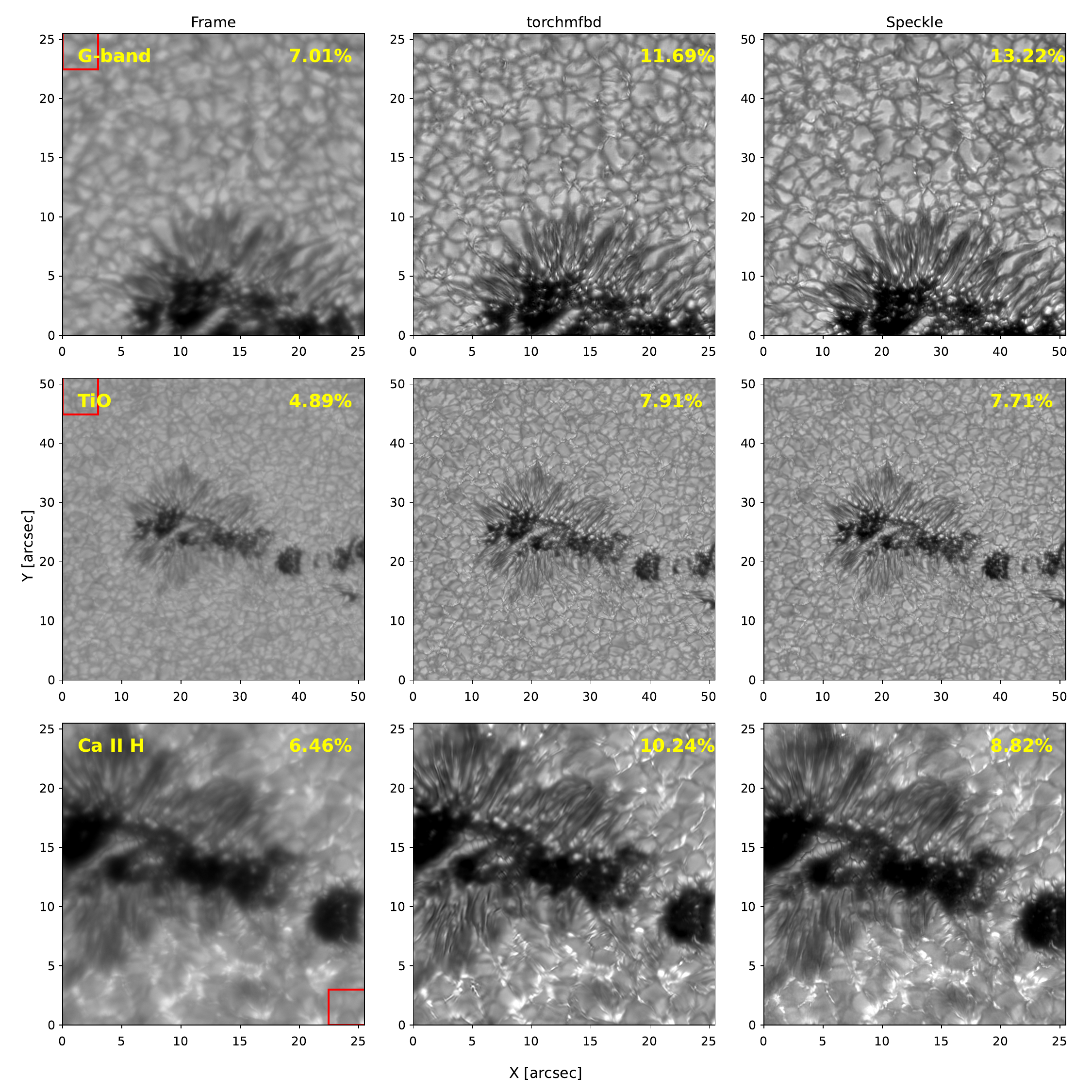}
  \caption{Reconstruction of HiFI data in the G-band, TiO band and Ca \textsc{ii} H. The first column corresponds
  to the best frame in the burst. The second column shows the reconstructed images with \torchmfbd\, while
  the last column displays the Speckle-restored images. The RMS contrasts given in the upper right corners are computed 
  in the small windows in the granulation marked in red.}
  \label{fig:hifi}
\end{figure*}

\subsection{IMaX}
Our first example is a simple phase-diversity (PD) image reconstruction using data from the IMaX 
instrument \citep{imax11} onboard the first flight of the \emph{Sunrise} balloon-borne observatory 
\citep{sunrise08,2010ApJ...723L.127S}. IMaX was a Fabry--P\'erot interferometer with polarimetric 
capabilities that observed the Fe~\textsc{i} line at 5250.2~\AA. The field of view was $46''\times46''$, 
with a spatial resolution of $0\farcs15$--$0\farcs18$ and a pixel size of $0\farcs055$ in two CCDs 
separated by a polarizing beamsplitter for dual-beam polarimetry.

The PD algorithm requires at least two images with the same unknown wavefront phase aberrations, but 
with additional, known differences in the aberrations. IMaX collected  PD data intermittently with a 
27~mm thick plate of fused silica inserted in front of one of the CCDs, producing the equivalent of 0.85~cm
displacement along the optical axis or  $1\lambda$ peak-to-peak defocus. The polarimetric data were 
then deconvolved with the PSF measured from the PD data nearest in time. This was feasible as the main 
sources of aberrations were due to fixed focusing of the cameras (0.5~mm), astigmatism in the etalons, 
and slowly evolving mechanical changes in the optics. With \emph{Sunrise} at stratospheric altitude, 
the telescope was well above the turbulence in the troposphere.

The focused and defocused frames for one such PD data set, with  quiet Sun in the FoV, are shown in the 
upper row of Fig.~\ref{fig:imax}. The lower left panel shows the reconstruction with \torchmfbd\ using 
a KL basis with 44 modes, while the lower right panel displays the reconstruction from the IMaX data 
release \citep{2010ApJ...723L.127S}. Assuming isoplanatic aberrations, the released  
reconstruction was made by deconvolving the data using a PSF based on an average of multiple wavefronts 
fitted with 45 Zernike modes. Those wavefronts are from PD processing of 30 PD image pairs 
in 10$\times$10 overlapping 128$\times$128-pixels patches over the full FoV. They used the PD 
algorithm of \citet{lofdahl_scharmer94}, ported to IDL by \citet{2003ASPC..307..137B}.
Also assuming isoplanatism, in the \torchmfbd{} restoration the whole FoV was processed in a single inversion.


Visually, the \torchmfbd{} restoration in Fig.~\ref{fig:imax} is sharper than the 
released restoration. This impression is consistent with the power spectrum in 
Fig.~\ref{fig:power}. 
The \torchmfbd{} image is restored to almost twice the 
resolution (0.2 vs 0.1 pix$^{-1}$) and also has slightly greater power in the 
0.02--0.1~pix$^{-1}$ range. The noise cutoff of the released image seems too low, 
considering where the noise dominates in the power spectrum of the raw data. Whether the power 
difference is due to the choice of basis functions or under-correction in the released 
image (because an average of many similar wavefronts is bound to be lower in RMS than 
any one of the wavefronts) is not clear.  The RMS contrast of the released
restoration (11.14\%) is lower than what is expected from 3D MHD simulations of 
the solar atmosphere. \citet{2010A&A...521A..68S} report 14.8\% RMS granulation 
contrast in the nearby 538~nm continuum ($\mu=1$). With negligible turbulence 
we are not affected by the uncorrected high-order modes that reduce the contrast 
and interfere with the wavefront fits for data from ground-based telescopes.  
High-order polishing (ripple) aberrations are estimated by \citet{2009PhDT........78V} 
to reduce the contrast by 5\% of the real value, which would give 14.06\% for 538~nm, 
still a bit above the \torchmfbd{} restored RMS contrast of 13.10\%. A possible explanation
is that the ripple not only reduces the contrast but also interferes with the fitting of a 
finite number of modes.

\subsection{CRISP} 
\label{sec:CRISP}

 Figure \ref{fig:crisp} shows the reconstruction of two CRISP observations in the \ion{Ca}{ii} 8542~\AA\ spectral line. In the two CRISP datasets, there are 12 exposures for each wavelength position and polarization state. The seeing-induced wavefront aberrations are approximately frozen with an exposure time of 17~ms. The cadences of the full spectropolarimetric scans were 20--30~s but for the restorations shown here, we use only the 12 exposures collected in 1.2~s of a single NB channel, together with the matching WB images. The WB images were collected through the CRISP prefilter with a 0.83~nm full-width at half-maximum (FWHM) wide passband covering the core and inner wings of the line but not the surrounding continuum \citep{2021A&A...653A..68L}. Only one of the four polarization states cycled during the data collection was selected. The WB burst
is destretched using the frame with the largest contrast as reference. The inferred optical flow is
then applied to the images of the NB channel.

The first two columns correspond to a $20''\times 20''$ sub-field of the entire FoV of $55''\times 55''$, showing the 
active region AR12767 on July 27, 2020 located at heliocentric 
coordinates $(-20'', -416'')$. These data were briefly described by \mbox{\citet{2022ApJ...928..101V}}. The NB channel of CRISP was tuned to 65~m\AA\ to the red of the core of the line, with a width of 105~m\AA{} FWHM. The seeing was fair, with $r_0$ from 9~cm to 
almost 16~cm (as measured at 500~nm with the AO wavefront sensor) during the 20~s scan and $\sim$12~cm when this particular subset was collected.
The rightmost two columns display a quiet Sun observation at disk center acquired on Aug 1, 2019 with a cadence of 31~s. The NB image also
corresponds to a displacement of 65~m\AA\ to the red of the core of the line. This is a subset of the data described and analyzed by \citet{2023A&A...672A.141E}. The seeing was excellent with $r_0$  between 15 and 20~cm throughout the scan.

MOMFBD without PD requires variety in the wavefronts. We would expect more of that and therefore a better restoration for a whole scan, as the data are regularly processed in the SSTRED pipeline. There is a trade-off however, in that the photosphere has time to evolve during a scan and this violates the assumption of a common object. In particular for the WB covering the entire scan but rapid motion in the chromosphere can challenge this assumption even in the short observations of the NB states \citep{2006ApJ...648L..67V}.

\begin{figure}
  \centering
  \sidecaption
  \includegraphics[width=\columnwidth]{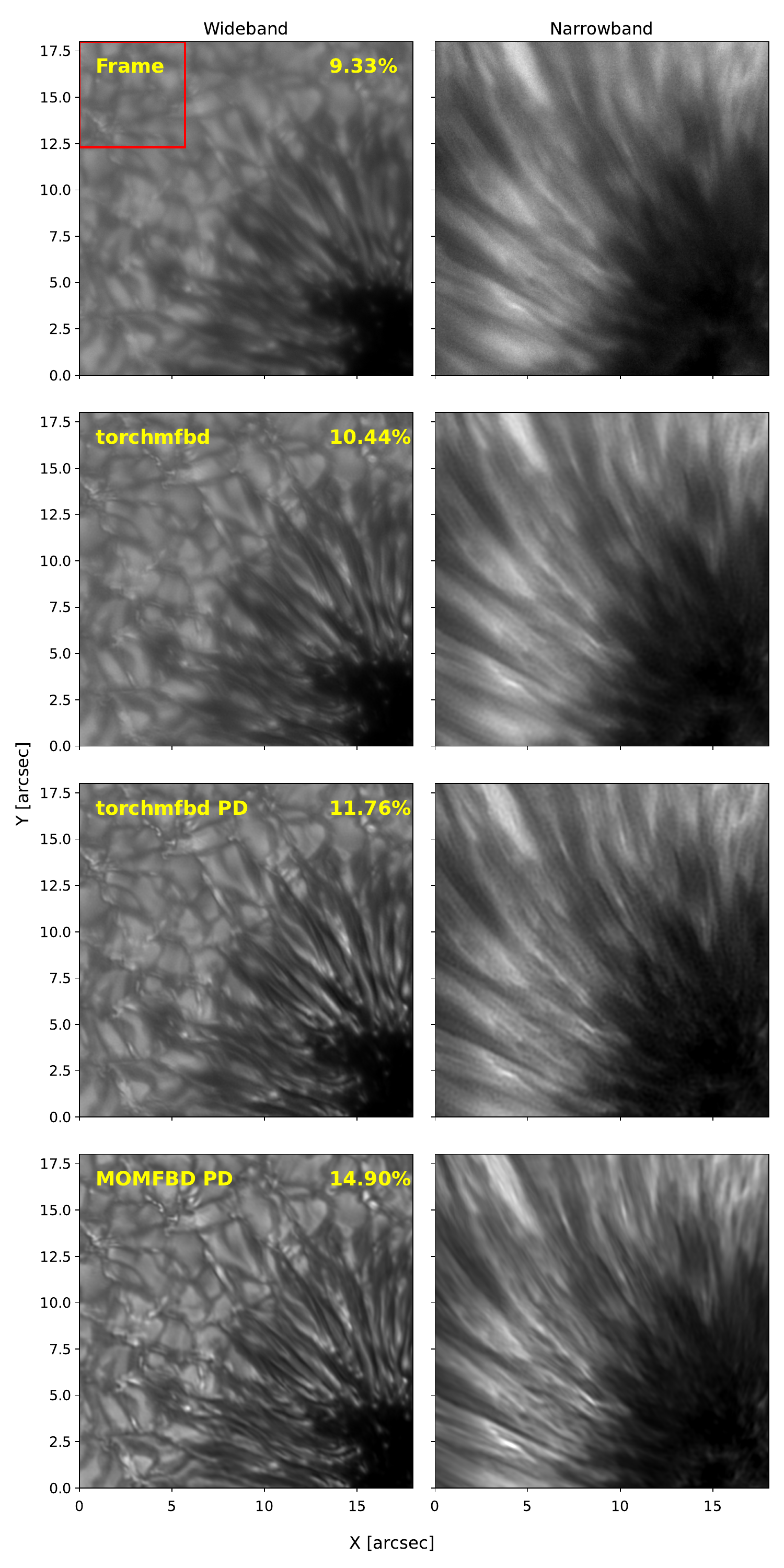}
  \caption{Examples of image reconstructions for data obtained in the Ca \textsc{ii} H line with CHROMIS. The 
  contrasts quoted in the panels
  were measured within the red rectangles.}
  \label{fig:chromis}
\end{figure}

The upper row of Fig.~\ref{fig:crisp} displays the best frame of the WB and NB channels, selected as 
the one with the highest RMS contrast. The second row shows the WB and NB images, reconstructed 
with \torchmfbd{} using the KL basis with 54 modes. The reconstruction was carried out in a 
NVIDIA RTX 4090 GPU with 24~GB of memory. The optimization process takes around 5~s for a FoV of 
512$\times$512 pixels using patches of 92$\times$92 pixels (or a square of side 5\farcs4). The number of modes and 
patch size are chosen so that they are as close as possible to the standard setting in the SSTRED data 
pipeline \citep{2021A&A...653A..68L}. The patches
are extracted every 40 pixels. Given the amount of data and memory available in the GPU, the reconstruction can be done in a single 
step using parallelization.
For comparison, the third row also shows the reconstruction carried out with \momfbd{} using the 
same number of KL modes. In each column, all images are shown in the same intensity scale.

The \torchmfbd\ reconstructions are visually similar to those of \momfbd{}, though with slightly 
higher contrasts. For a more quantitative comparison, the lower leftmost panel of Fig.~\ref{fig:power} 
displays the WB power spectrum of the quiet Sun observation. Assuming the power law part of the power
spectrum can be extrapolated below the noise level of individual frames, it is reasonable to expect 
the \torchmfbd{} restored power to approach the diffraction limit, as the noise level decreases 
with the number of frames.
However, the dataset is small and collected in a very short time, so there is some chance that we are 
simply looking at the mosaicking artifacts at the highest frequencies. The \momfbd{} restoration is by default limited to 90\% of the 
diffraction limit. 

The power in the signal-dominated frequencies is slightly greater with \torchmfbd, consistent with the higher contrast. 
We have not investigated the reason for the deviation between both reconstructions, but it probably is a
consequence of the implementation differences between both codes, that might affect both 
the convergence of the iterative fitting and 
the noise filter construction (which is done 
at the patch size). The main message here is that \torchmfbd{} appears to work at least as well as the \momfbd{} code.

The bottom row of Fig.~\ref{fig:crisp} displays the results applying the still experimental spatially variant 
blind deconvolution using the NMF dictionary. This FoV can be reconstructed in around 10~s using the GPU mentioned above. 
The results are encouraging and they motivate the necessity of a deeper study of image reconstruction using NMF, something 
that we plan to carry out in the near future. For the purpose of reproducibility, the reader can find the hyperparameters 
of the configuration files and the scripts in the code repository.

\subsection{HiFI}
An example from GREGOR, using observations from three different filters of the High-resolution Fast 
Imager \citep[HiFI,][]{hifi}, is shown in Fig.~\ref{fig:hifi}. This is an example of single-object multiframe blind deconvolution (MFBD). The observations of AR13378 were carried out 
on July 21, 2023. The single exposure times ranged from 2.5 to 9.0\,ms, depending on the wavelength. Sets of 500 
images were acquired with a cadence of approximately 11\,s. The images were then reduced and calibrated using 
the sTools pipeline \citep{stools}. During data reduction, only the best 100 images from each set were selected. 
Finally, these selected images were processed using KISIP \citep{KISIP} to generate a single Speckle-restored image. 
The \torchmfbd\ reconstructions were obtained using 135 KL modes, using burst of
100 images with patches of size $96\times 96$ with a step of 50 pixels for the observations
of the TiO-band, $64\times 64$ with a step of 20 pixels for Ca~\textsc{ii}~H and $32\times 32$ with a step of 10 pixels for
the G-band. These large overlaps serve to reduce the merging artifacts when recomposing the FoV, but they can surely be  
optimized to reduce the computing requirements. The 
\torchmfbd\ reconstructions are visually very similar to the Speckle-restored images, though providing slightly higher
granulation contrasts. Both speckle and MFBD struggle to reach the expected
contrast of 21.5\% in the G-band \citep{2007ApJ...668..586U}, speckle methods probably because they
are sensitive to the calibration of the AO correction and MFBD methods because they do not
correct for very high-order modes in the wavefront.
More quantitatively, we show the power spectra on the upper row of Fig.~\ref{fig:power}, demonstrating that
\torchmfbd\ is able to recover very similar power in all spatial frequencies when compared with the Speckle restorations. The 
reconstructions push the effective resolution of the observations close to the diffraction limit, indicated
by the vertical line.

\subsection{CHROMIS}

The last example is CHROMIS data obtained in the \ion{Ca}{ii}~K
spectral line. These data are part of the same observation as the
active region CRISP dataset described in Sect.~\ref{sec:CRISP} and
share the same seeing conditions. The whole scan has a cadence of 6~s,
15 frames per wavelength position with an integration time of 12.5~ms.
Like for the CRISP data, we MOMFBD-processed NB data for a single
tuning, $65$~m\AA{} from the line core, together with the
simultaneous WB frames at 3950~\AA{}. CHROMIS is equipped with a
defocused WB camera, and we processed the data both with and without the
simultaneous PD frames. The 15 exposures were collected in less than
0.2~s while $r_0$ was $\sim$11~cm. 

Given the large FoV observed with CHROMIS, we show in
Fig.~\ref{fig:chromis} the results for a subarea of
$20\arcsec\times20\arcsec$. The reconstruction without PD (second row)
provides a slight visual improvement when compared with the best raw frame
(first row). One cannot expect much improvement with data collected during such
a short time. Better results are obtained when the PD channel is
included (third row). All \torchmfbd\ calculations have been performed with patches of 
size $128 \times 128$~pixels with 54 KL modes. The aim was to reconstruct
in conditions as close as possible to those of the \momfbd{} reconstructions (last row),
which have been computed with patches of size 136$\times$136~pixels with 54 KL modes. 

The contrast in the quiet Sun region of the observed FoV increases monotonically once
the reconstruction is performed, with our \torchmfbd\ results slightly lagging behind 
those of \momfbd{}. The general appearance of the image and, especially, some small details appear better
resolved (the inner penumbra in the WB image and some superpenumbral
filaments in the NB image) with \momfbd{} than with \torchmfbd. 
The quantitative comparison is displayed in the central lower panel of
Fig.~\ref{fig:power}. The resolution is far below the diffraction
limit because better seeing is required for excellent observations at 400~nm.
However, the power at all frequencies is very similar between both codes
when including PD.

\section{Conclusions}
\label{sec:conclusions}

We have introduced \torchmfbd, a modern and versatile implementation of the 
MOMFBD method. This open-source code is designed to address the challenges of post-facto image 
restoration for ground-based astronomical observations. The use of the \texttt{PyTorch} library allows \torchmfbd\ to be
flexible and take advantage of GPU acceleration, significantly reducing both development and computation times.



In \torchmfbd{} we have implemented most of the MOMFBD algorithm of the \momfbd{} code and also introduce a few innovations. New are:
\begin{enumerate}
\item the spatially variant convolution of Sect.~\ref{sec:spat-vari-conv}
\item the PSF parameterization of Sect.~\ref{sec:PSF}
\item the smoothness regularization of Sect.~(\ref{sec:smoothness})
\item the sparsity regularization of Sect.~\ref{sec:sparsity}
\item the destretching of input images of Sect.~\ref{sec:destretching}
\item the GPU acceleration described in Sect.~\ref{sec:timing-memory-cons}
\end{enumerate}
We use some of them for the demonstrations in Sect.~\ref{sec:results}. A thorough evaluation of how to 
best combine them has to be done separately for different kinds of data and is outside the scope of this paper.
To be implemented soon are the linear equality constraints and the
proximal projection algorithm mentioned in Sect.~\ref{sec:optim-paral}.

\torchmfbd\ is publicly available, along with comprehensive documentation and examples to facilitate its adoption
by researchers. A modern, flexible, and efficient framework for MOMFBD, \torchmfbd\ provides
researchers with an extensible code for testing and implementing new ideas in the field of image restoration 
with methods based on multi-frame blind deconvolution.

\begin{acknowledgements}
AAR, SEP and CK acknowledge funding from the Agencia Estatal de Investigación del Ministerio de Ciencia, Innovación y Universidades (MCIU/AEI) under grant 
``Polarimetric Inference of Magnetic Fields'' and the European Regional Development Fund (ERDF) with reference PID2022-136563NB-I00/10.13039/501100011033.
SEP acknowledges the funding received from the European Research Council (ERC) under the European Union’s Horizon 2020 research and innovation 
programme (Advanced grant agreement No 742265). This research is supported by the Research Council of Norway, project number 325491, and through its 
Centers of Excellence scheme, project number 262622.
CK acknowledges grant RYC2022-037660-I funded by MCIN/AEI/10.13039/501100011033 and by "ESF Investing in your future".
MGL acknowledges grant 2023-00169 for EST-related research from the Swedish Research Council. 
The computations were partly performed on hardware of the Institute for Solar Physics (ISP), financed 
by the same grant. The authors acknowledge C. Sasso, who was the PI of the GREGOR campaign.
The authors thankfully acknowledge the technical expertise and assistance provided by the Spanish Supercomputing Network 
(Red Española de Supercomputación), as well as the computer resources used: the LaPalma Supercomputer, located at the Instituto de 
Astrof\'{\i}sica de Canarias. This paper made use of the IAC HTCondor facility (\texttt{http://research.cs.wisc.edu/htcondor/}), 
partly funded by the Ministry of Economy and Competitiveness with FEDER funds, code IACA13-3E-2493.
The Swedish 1-m Solar Telescope is operated on the island of La Palma by the Institute for Solar 
Physics of Stockholm University in the Spanish Observatorio del Roque de los Muchachos of the Instituto de Astrofísica de Canarias. 
The Swedish 1-m Solar Telescope, SST, is co-funded by the Swedish Research Council as a national research infrastructure 
(registration number 4.3-2021-00169). This research has made use of NASA's Astrophysics Data System Bibliographic Services.
We acknowledge the community effort devoted to the development of the following open-source packages that were
used in this work: \texttt{numpy} \citep[\texttt{numpy.org},][]{numpy20}, 
\texttt{matplotlib} \citep[\texttt{matplotlib.org},][]{matplotlib}, \texttt{PyTorch} 
\citep[\texttt{pytorch.org},][]{pytorch19}, \texttt{scipy} \citep[\texttt{scipy.org},][]{2020SciPy-NMeth} and \texttt{scikit-learn} \citep[\texttt{scikit-learn.org},][]{scikit-learn}.
\end{acknowledgements}

%
%

\end{document}